\documentclass[a4paper,11pt]{article}
\pdfoutput=1 

\usepackage{jcappub} 

\usepackage[T1]{fontenc} 

\title{\boldmath Large-Volume Centimeter-Wave Cavities for Axion Searches}

\author[a,b]{Chao-Lin Kuo}

\affiliation[a]{Kavli Institute of Particle Astrophysics and Cosmology \& Physics Department,
Stanford University, Stanford, CA 94305, USA}
\affiliation[b]{SLAC National Accelerator Laboratory, Menlo Park, CA 94025, USA}

\emailAdd{clkuo@stanford.edu}

\abstract{The scan rate of an axion haloscope is proportional to the square of the cavity volume. In this paper, a new class of thin-shell cavities are proposed to search for axionic dark matter. These cavities feature active volume much larger ($>20\times$) than that of a conventional cylindrical haloscope, comparable quality factor $Q$, and a similar frequency tuning range. Full 3D numerical finite-element analyses have been used to show that the TM$_{010}$ eigenmodes are singly polarized throughout the volume of the cavity and can facilitate axion-photon conversion in uniform magnetic field produced by a superconducting solenoid. To mitigate spurious mode crowding and volume losses due to localization, a pre-amplification binary summing network will be used for coupling. Because of the favorable frequency-scaling, the new cavities are most suitable for centimeter-wavelength ($\sim$ 10$-$100 GHz), corresponding to the promising post-inflation axion production window. In this frequency range, the tight machining tolerances required for high-$Q$ thin-shell cavities are achievable with standard machining techniques for near-infrared mirrors. }

\begin{document}
\maketitle
\flushbottom

\section{Introduction}

Axion, the light scalar particle that was proposed to solve the strong {\em CP} problem in QCD (quantum chromodynamics), is an excellent candidate for dark matter \cite{pq,weinberg,wilczek}.  Resonant conversion to microwave photons in the presence of strong magnetic field is a mature method to search for relic QCD axions \cite{sikivie1}. This method, known as the axion haloscope, has produced the most sensitive limits on axion coupling strength between $0.5-7$ GHz, corresponding to an axion mass of $2-30\;\mu$eV \cite{admx,haystac,haystac_result,admx_side}. 

The resonator design adopted by these haloscope experiments consists of a cylindrical cavity with one or multiple vertical rod(s). The frequency tuning is achieved by rotating the off-axis rods and changing the physical configuration of the cavity.  
The scan rate $d\nu/dt$ of an amplifier noise-limited axion haloscope is proportional to $B_0^4V^2 Q$, where $B_0$ is the applied static magnetic field, $V$ is the {\em active} volume of the cavity mode coupled to the magnetic field, and $Q$ is the quality factor of the resonator \cite{sikivie1,graham,irastorza}. Since $V$ scales as $\nu^{-3}$, at 10 GHz the scan rate is already down by six orders of magnitude compared to 1 GHz. This unfavorable scaling is exacerbated further by higher amplifier noise and lower quality factors achievable at higher frequencies. This represents substantial missed opportunities because an axion frequency $\sim $ 10 GHz corresponds to the important post-inflation axion generation mechanism, sometimes known as the classical window \cite{hertzberg,marsh,graham2}. 

In recent years, there have been an explosion of new ideas on axion searches, including many new concepts that are not relying on cavities \cite{plasma, dmradio, nmr, abra, lumped, madmax, irastorza}. Although this paper is focusing on the design of cavity resonators, it has certainly been partly inspired by these exciting innovations.  The proposed new architecture borrows many methods well known in the cosmic microwave background (CMB) community. It can work together with novel quantum techniques that reduce readout noise, by either operating JPAs (Josephson Parametric Amplifiers) in a squeezed setup to defeat quantum limit \cite{quantum1,quantum2}, or by using photon counting Qubits or Rydberg atoms to completely circumvent it \cite{qubit1,rydberg}. 

In the past, there have been a few exploratory studies on alternative cavity designs. The dominant idea is to combine data from multiple cavities or cavity cells to increase the effective volume, while fixing the resonant frequency in the desired range \cite{cell1,rades,organ}.  Implementations of these ideas often face significant challenges in synchronous tuning. Another idea is to employ TE modes in long rectangular cavities \cite{baker}. The main disadvantage of this approach is that the linear geometry leads to substantial difficulties in cryogenics and magnetic coupling. Their recent proposal is to break this cavity into multiple cells, essentially going back to the first idea. 

In this paper, a class of novel thin-shell cavities are proposed for axion searches in the few to tens of GHz (centimeter-wavelength) frequency range. The active volume of the new designs scales much more mildly, as $\propto \nu^{-1}$, or even weaker if a more elaborate variety can be realized. Unlike the linear resonators, this more compact design fits inside a reasonably-sized cryostat and eases difficulties in magnetic coupling.  Finite-element analyses (FEA) based on COMSOL-RF have been used to verify the resonant frequencies of the linearly polarized TM$_{010}$ mode, as well as the high quality factor of the new design. At 15 GHz ($\lambda=$ 2 cm), the active volume of a circular thin-shell design is more than 20 times that of a scaled HAYSTAC cylindrical cavity, corresponding to a $>$400$\times$ improvements in the scan rate. In principle, the active volume is only limited by practicalities, including fabrication tolerances, spatial localization of the desired mode, and mode crowding by other resonances. We propose to utilize polarization and symmetry properties to suppress coupling to undesired spurious modes. Specifically, a phase-matching binary summing network similar to the ones used successfully in millimeter-wave astronomy will be used to read the cavity signal from multiple polarization selective ports. 

The rest of the paper is organized as the following: Section 2 discusses the resonant properties of the thin-shell cavities, Section 3 discusses practical issues, including fabrication, and a mode-managing summing network, Section 4 covers aspects of magnetic coupling and frequency tuning. Section 5 discusses projected sensitivity to QCD axions and summarizes the paper. 

\section{Geometry and Properties of Large-Volume Resonators}

The axion field interacts with a static magnetic field and creates an effective oscillatory current $j_a$ that excites a cavity eigenmode by coupling to the $E$ field \cite{sikivie1,graham,irastorza}. Therefore, for a uniform external $B$ field the appropriate eigenmode should be linearly polarized. We start by reviewing resonant behaviors in a simple rectangular cavity with linear dimensions ($L, w, h$) in the ($x,y,z$) directions (FIG. 1).  The TE$_{lmn}$ modes have resonant frequencies given by

\begin{equation}
    \nu_{lmn}=\frac{c}{2}\sqrt{(l/L)^2+(m/w)^2+(n/h)^2},
\end{equation}

where $c$ is the speed of light and $(l,m,n)$ are non-negative integers \cite{balanis}. If the external magnetic field is applied in the $z$ direction (along $h$), the TE$_{110}$ mode is most suitable for axion searches because of its uniform polarization. In this case, the $E$ field distribution is sinusoidal across $L$ and $w$, and uniform across $h$. 
In \cite{baker}, the dimension $L$ is taken to be the full length (a few m) of an existing dipole magnet system at CERN. In that proposal, the height $h$ must fit inside a small gap between the dipole magnets. While the volume of this pencil beam cavity is indeed large for the frequency, the geometry leads to practical limitations \cite{rades}. 

In the proposed new design, $h$ and $L$ are {\em both} taken to be much larger than $w=\lambda/2$, and additionally the $L$ dimension is rolled up into a circle. The TE$_{110}$ mode is turned into a TM$_{010}$ mode living in a thin shell as the $x$ axis becomes $\phi$ and $y$ becomes $\rho$. The eigenmode can be seen as a $z-$polarized quasi-plane wave bouncing back and forth between the inner and outer walls, setting up a standing wave (FIG. 1, {\em right}). Analytic calculations verify that the existence of eigenmodes in a closed form, $E_z(\rho)=c_1 J_0(k\rho) + c_2 Y_0(k\rho)$, $E_\phi=E_\rho=0$.  The coefficients to the Bessel Functions of the first ($J_0$) and the second ($Y_0$)  kind are fixed by the boundary conditions requiring $E_z$ to vanish at the walls ($\rho=r$ and $\rho=r+w$). 

For this mode, the resonant frequency approaches a fixed value when $L$ and $h$ are taken to be much larger than $w\sim\lambda/2$, and the cavity still resonates at $c/\lambda$ as its volume ($V=Lwh$) increases (FIG. 2).  In Section 3 of the paper, several methods have been proposed to isolate the desired TM$_{010}$ mode out of a forest of spurious modes. While this is a challenging process for a very large cavity, it should be pointed out that there is a continuous spectrum of volume/sensitivity to be gained this way.  How much volume can eventually be realized in an experiment depends on fabrication tolerances achieved and how successful the mitigation schemes described in Section 3 turn out to be in practice. 
\begin{figure}[t]
\centering 

\includegraphics[width=4.5in]{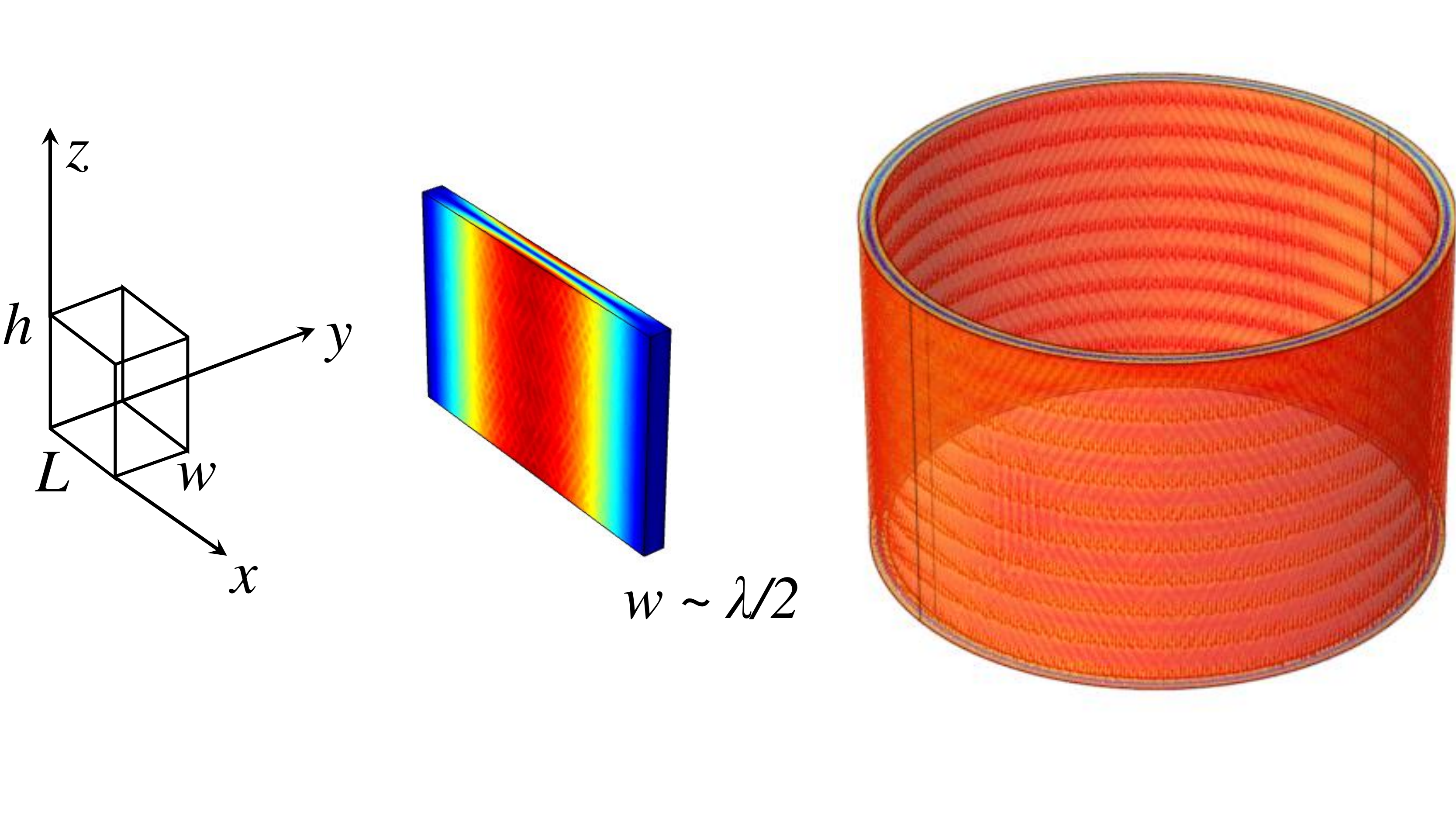}
\caption{\label{fig:epsart2} {\em Left:} A rectangular resonant cavity. This simple cavity supports a TE$_{110}$ mode that is polarized in the $z$ (vertical) directions. {\em Center:} A flattened rectangular cavity with extended dimensions in $L$ and $h$, and a smaller $w$. According to Eq. (1) the TE$_{110}$ mode (shown as surface current distribution) now resonates at $\sim c/2w$. {\em Right:} A cavity formed by rolling up the flattened rectangular cavity into a circular thin shell. Analytic calculations and FEA verify the intuitive guess that this shell cavity supports a $z-$polarized resonance. The surface current distribution calculated by COMSOL is shown.}
\end{figure}

One can also add $\lambda/4$-deep azimuthal corrugations \cite{balanis,dragone}, widely used in radio astronomy and accelerator cavities, to the top and bottom of the thin shell cavity. With corrugations the uniform $z-$distribution becomes sinusoidal, while the mode still maintains overall linear polarization throughout the shell. Such structure forces $E_z$ to vanish at the top and bottom boundaries (FIG. 2).  This property can facilitate frequency tuning via a mechanical means, as will be elaborated in section 4. This is essentially a corrugated rectangular cavity (that supports a high-$Q$ HE$_{111}$ mode) squeezed in the propagation direction and rolled up into a circle in one of the two aperture dimensions \cite{dragone}.  
Basic scaling relations of resonant frequencies for both the corrugated and uncorrugated circular shell cavities are verified using FEA (FIG.3). The nominal resonant frequency is chosen to be 15 GHz throughout the paper unless otherwise stated.

\begin{figure}[t]
\centering 

\includegraphics[width=4.in]{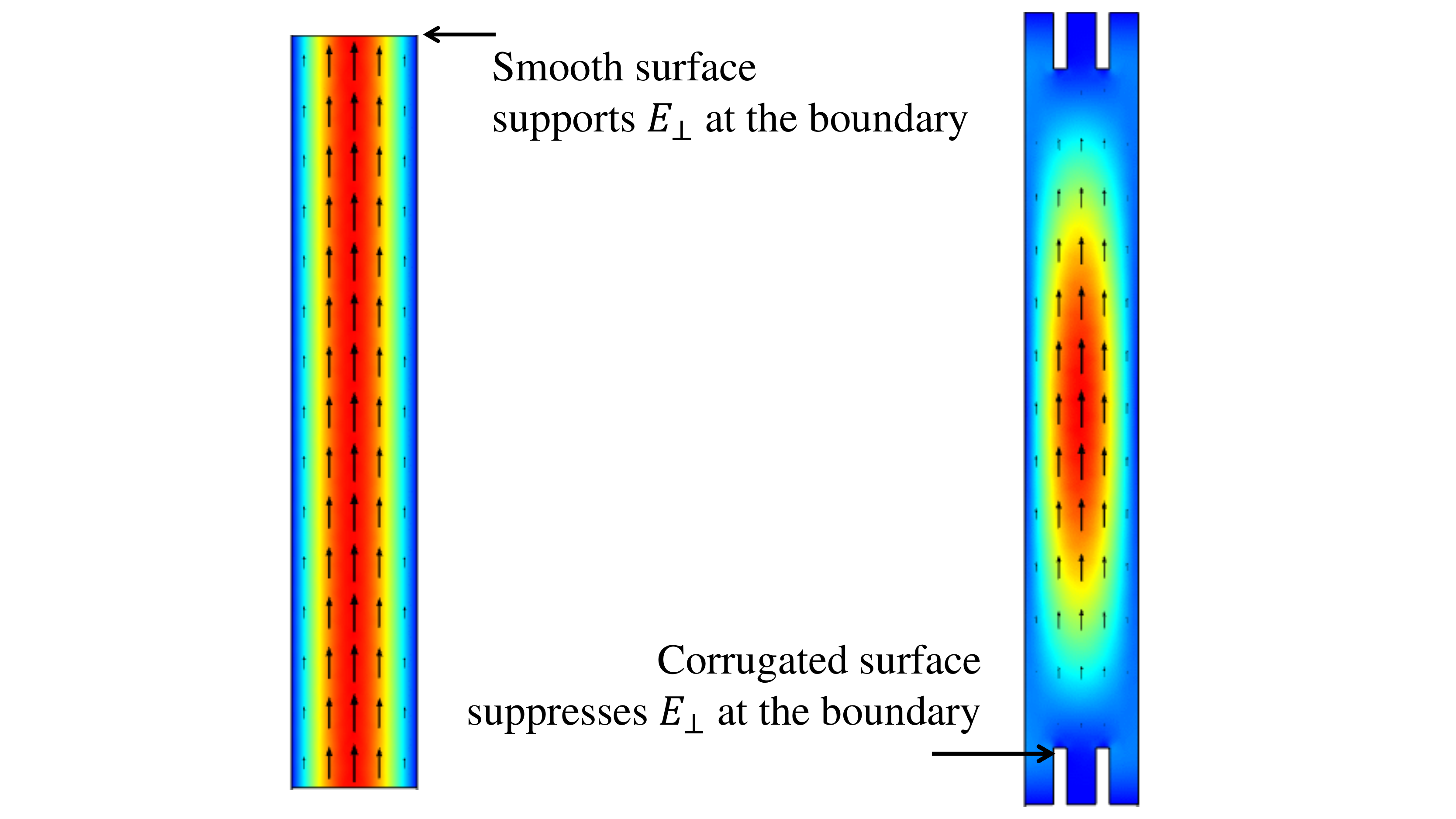}
\caption{Sub-wavelength corrugations ($\lambda/4$) force $E_\perp$ to vanish at the boundaries. {\em Left:} Field distribution in an uncorrugated circular-shell cavity; {\em Right:} Field distribution in a corrugated cavity.  As is discussed in later sections, the suppression of fields at the boundaries helps facilitate frequency tuning. 
}
\end{figure}

\begin{figure}[t]
\centering 
\includegraphics[width=3.3in]{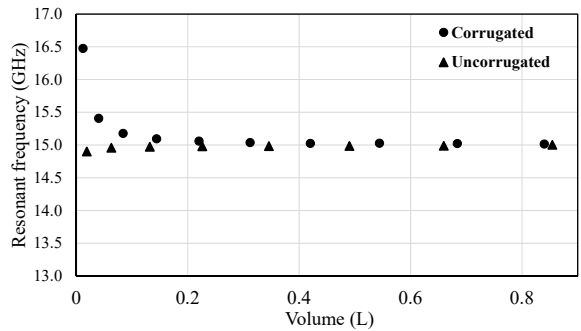}
\caption{As the volume increases, the resonant frequency of the singly polarized TM$_{010}$ modes approaches a constant value determined by the wall separation $w$ in a circular shell resonator. Two sets of symbols represent results with/without corrugations on the top and bottom.  The calculations have been done with COMSOL-RF, for the case of $w=1$ cm, $r=1,2,..,10$ cm, and $h=2r$.}
\end{figure}

\begin{figure}[t]
\centering 
\includegraphics[width=3.3in]{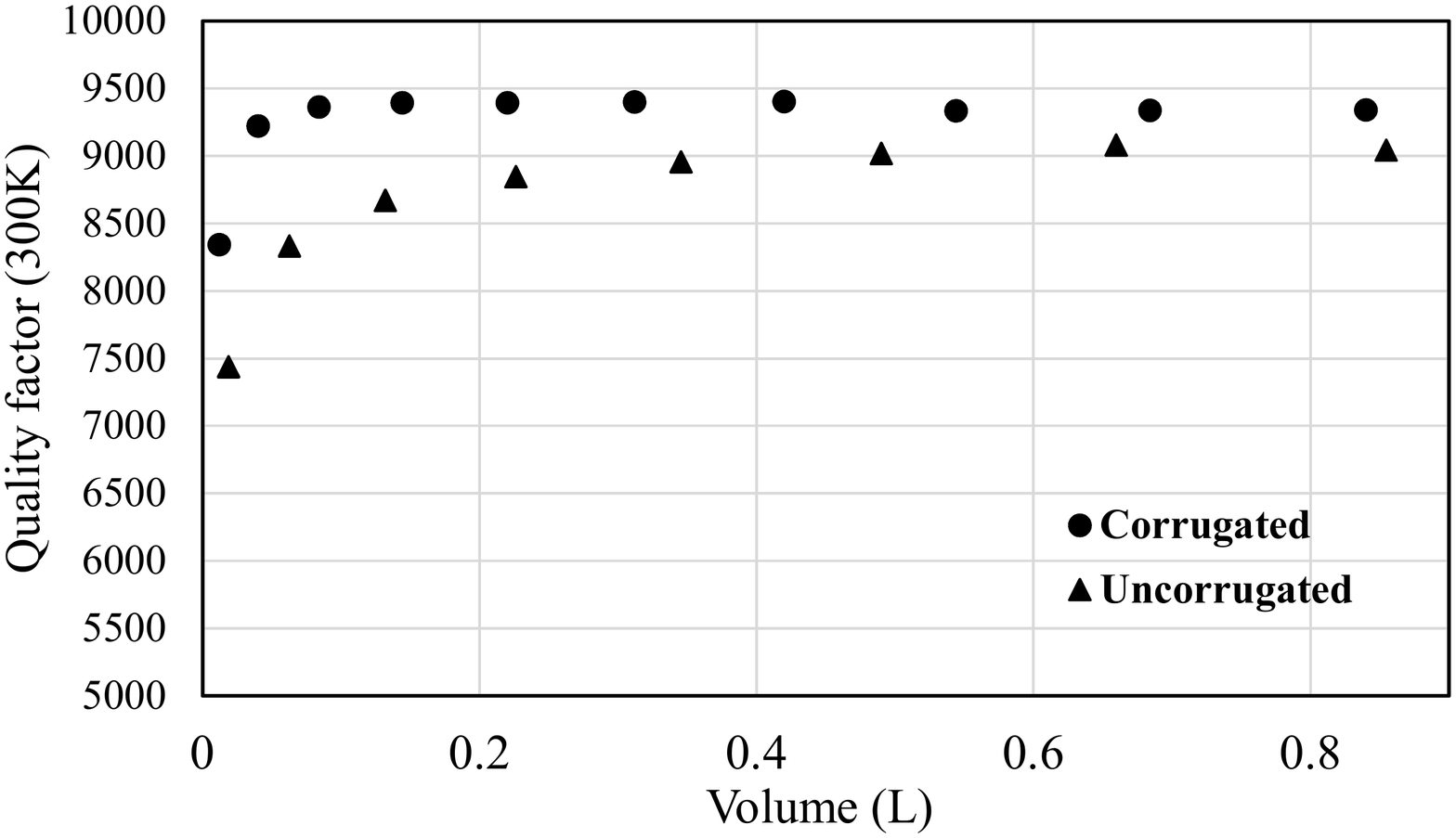}
\caption{These FEA calculations directly show that the quality factor stays constant as the cavity volume increases in a circular shell resonator, confirming the scaling arguments given in the main text. The calculations have been done for the case of $w=1$ cm, $r=1,2,..,10$ cm, and $h=2r$.}
\end{figure}

A representative large volume circular shell cavity that resonates at this frequency has an inner radius $r$ of 15 cm, an outer radius of $r+\lambda/2=$ 16 cm, and a height $h$ of 20 cm (FIG. 5, {\em left}). COMSOL-based 3D FEA's confirm the existence of a linearly polarized TM$_{010}$ mode sitting at 15.0 GHz in the new cavity.  Incidentally, the new geometry can also be obtained by increasing {\em both} the inner radius $r$ of a legacy cylindrical cavity and the radius of its tuning rod, while maintaining the gap to be $\sim\lambda/2$. To compare the active volume of a legacy resonator with the new thin shell design, we can scale down the HAYSTAC cavity \cite{brubaker} by a factor of $2.57$ in all dimensions (FIG. 5, {\em center top}), so that the highest resonant frequency is also at 15 GHz (confirmed by FEA). The two cavities have nearly identical form factors for the TM$_{010}$ mode.  We see that the new design has a volume that is $20.9$ times larger than the scaled HAYSTAC resonator.  The larger volume represents a very significant enhancement in the scan rate, since $d\nu/dt$ is proportional to $V^2$.  

The form factor for each eigenmode, usually denoted by $C_{lmn}$ in the axion literature, measures the overlap of $E$ and $B$ within the cavity.  If the external magnetic field is uniform in $z$, the form factor is given by $(\int E_z dV)^2/(V\int E^2 dV)$.  Since $C_{010}<1$, the effective volume of the cavity $C_{010}V$ is generally reduced to be smaller than $V$.  The form factor can be calculated either numerically using FEA results, or in the case of circular thin-shell cavities, analytically using the asymptotic forms for the Bessel functions: $J_0(x)\sim \cos(x-\pi/4)/\sqrt{x}$ and $Y_0(x)\sim \sin(x-\pi/4)/\sqrt{x}$.  For uncorrugated thin-shell cavities, the form factor works out to be $C_{010}\approx 0.811$, which asymptotically approaches $8/\pi^2$ when $w/r \rightarrow 0$ and the $E_z$ become purely sinusoidal in $\rho$.  For corrugated shell cavities, the extra sinusoidal dependence of $E_z$ in the $z$ direction reduces $C_{010}$ to $\approx 0.657$.  In other words, the effective volume in a corrugated cavity is reduced by about $20\%$ compared to an uncorrugated cavity with the same dimensions.  These numbers should still be very good approximations for general brain-type cavities, as long as the dimensions in $\rho$ and $z$ are much larger than $w$, when the conceptual picture of a $z-$polarized quasi-plane wave bouncing back-and-forth between the inner and outer walls becomes an accurate depiction.

Another important quantity, the quality factor $Q$, remains about the same even as the transverse dimensions are inflated. It is easy to see that in the new thin-shell design the surface current loss is dominated by the inner and outer walls (FIG. 1), which have a total area $A$ proportional to the volume ($V\sim Aw$). Therefore, the $Q$-factor, which is proportional the ratio of the volume and the surface loss \cite{balanis}, largely stays the same as $r$ and $h$ (and therefore $V$) increase indefinitely.  Direct numerical calculations confirm that in designs where $r\sim h \gg w $, the $Q$-factors approach a constant value modestly higher than that of a single-rod cylindrical cavity (FIG. 4).

\begin{figure}[b]
\centering 

\includegraphics[width=4.0in]{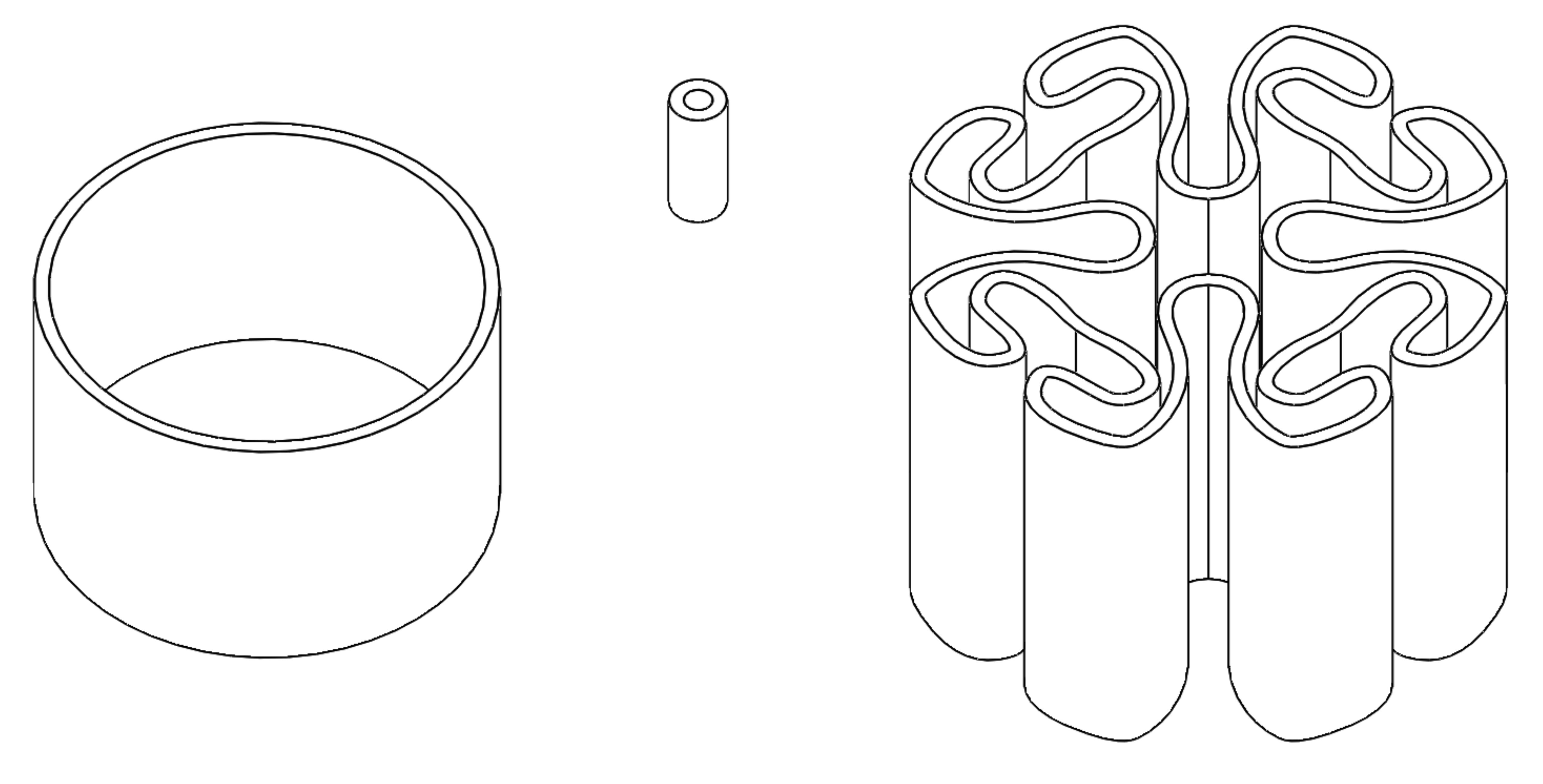}
\caption{\label{fig:epsart3} {\em Left:} A circular thin shell cavity ($r$=15 cm, $h$=20 cm, $w$=1 cm) is 20.9 times larger in volume than a conventional cylindrical cavity at the same frequency ({\em center top}), with a similar quality factor. This quasi-planar standing wave concept can be exploited further to improve the scan rate by an even larger factor, if the ``brain'' type cavity ({\em right}) can be fabricated with sufficient tolerances.}
\end{figure}

The unlimited transverse dimensions of a quasi-plane wave can be exploited further, by going into the radial dimension with structures that resemble the cerebral cortex (FIG. 5, {\em right}). The volume of this cavity could be nearly independent of the frequency as the number of folds increases. As long as the inter-wall distance is fixed at $\lambda/2$=1 cm, this cavity should have a resonance at 15 GHz and similar $Q$ based on the arguments given above. The particular cavity shown has an active volume that is more than 100 times larger than a scaled cylindrical cavity. Such a ``brain'' cavity would be even more difficult to meet the tolerance requirements, although the factor of $> 10^4$ improvement in scan speed should serve as a strong motivation to try. A cavity with this geometry is difficult even to simulate with FEA, however analytic treatments under WKB assumptions ($\partial_\perp \psi \gg \partial_\parallel \psi$, where $\psi$ is any field quantity) should be quite straightforward.  

The varying radius of curvature of the walls in a brain cavity slightly shifts the resonant frequency around. If not compensated it would lead to a reduction in the $Q$-factor. The size of this effect can be readily calculated for uncorrugated cavities. In a thin-shell cavity, the resonant frequency is given by $c/2w$ to the leading order, as mentioned earlier. For a circular shell, the next order term can be shown to be $-(c/2w)\cdot (w/r)^2/8\pi^2$ using Stokes' asymptotic expression for the roots of Bessel Function $J_0(x)$.  The size of the correction is on the order of $10^{-4}$ for $w/r\sim 0.1$.  As discussed in the next section, such tolerance level is challenging but achievable with modern fabrication techniques. 

It is rather straightforward to see that the proposed topological (FIG. 1) and homeomorphic (FIG. 5) gymnastics would lead to high-$Q$ cavities with indefinitely large active volume, because a linearly polarized quasi-plane wave has two extended transverse dimensions. It is also apparent that the resonant frequency of a single-rod cylinder is determined by the gap between the tuning rod and the inner wall ($\sim\lambda/2$). Therefore, a large-volume resonator is available simply by increasing the diameter of the rod and the cavity ID {\em together}. It does not matter that the cavity dimensions are much larger than the Compton wavelength of the axion. 

Obvious it may seem, to our knowledge no cavity geometries similar to those depicted in FIG. 5 have been proposed for axion searches in the past. Nevertheless, the main challenges for overmoded cavities are well known in the community, {\em e.g.}, see \cite{irastorza} for a recent comprehensive review. They are (1) fabricating the cavities with sufficient machining tolerances; (2) controlling the spatial distribution of the desired eigenmodes over a volume much larger than $\lambda^3$; and (3) mitigating mode crowding caused by the large volume. 

Despite the higher mode density, it is worth pointing out that modern microwave equipment and DAQ systems have enough bandwidth and resolution to record and resolve all the potential resonances. The analysis can then be performed offline with specialized algorithms. As long as the mode spacing is larger than the resonance bandwidth, the presence of other resonances does not hurt the search sensitivity.  The exception is at resonance crossing, when the TM$_{010}$ mode and a spurious mode ({\em e.g.} a TEM line) are in superposition.  When this happens, to linear order the induced axion signal will stay unaffected but the thermal noise from the spurious mode will degrade the signal-to-noise ratio.  The next section discusses methods making use of symmetry and polarization properties to reduce contributions from spurious modes.




\section{Fabrication, Coupling, and Other Implementation Issues}


The resonant frequency of a vertically polarized cavity mode is determined by the distance between the walls.  Random variations in the inter-wall distance broaden the resonant line and lower the quality factor accordingly. Therefore, the distance between the walls must be maintained at $\lambda/2$ to a precision of $\sim\lambda/2Q$ throughout the cavity shell.  If the machining tolerance is not met, more spurious modes with poorly understood behaviors can appear and hybridize with the TM$_{010}$ mode.  At $15$ GHz and $Q\sim 1.5\times 10^4$, this corresponds to a machining tolerance of $\sim 0.7 \;\mu$m. This is about the required precision for near-infrared mirrors and are routinely achieved using single-point diamond turning on stress-free metal blocks. The metal ({\em e.g.} nickel) parts can then be plated with high purity copper. Precision cylindrical grinding may be able to achieve the required tolerances for the circular thin shell. 

Once fabricated, the spatial distribution of the cavity TM$_{010}$ mode can be systematically surveyed using the ingenious ``bead mapping'' measurements described in \cite{kenany} and \cite{rapidis19}, in which a dielectric bead suspended by Kevlar threads is raster-scanned in the cavity space and the corresponding frequency shifts in the resonance is used to measure the $E$ field strength. One can imagine constructing a circular thin-shell resonator (or even a brain type resonator) out of multiple adjustable, diamond-turned panels and use the bead-mapping data to adjust the locations of the panels. Whether this extra complexity is needed or beneficial depends on the fabrication tolerances achieved. 

The large active area inevitably leads to mode crowding because it supports a large number of resonances. All the TM$_{l1n}$ modes resonate at only slightly higher frequencies than that of TM$_{010}$. Furthermore, for each TM$_{l1n}$ mode, there is a corresponding TE$_{l1n}$ at a similar frequency. On top of all this, a very large number of TEM-like modes can exist around the frequency of the TM$_{010}$ mode. 

We plan to utilize polarization properties and pre-amplification phase cancellation to mitigate undesired mode crowding and spatial localization caused by the remaining fabrication errors and misalignments. Fortunately, the polarization and symmetry properties of these spurious modes can all be simulated and analyzed. Using FEA, we observe that of these modes only the TM modes are polarized in the vertical ($z$) direction. Therefore, the first line of defense against resonance crowding by TEM-like ($\rho$-polarized) and TE-like ($\phi$-polarized) modes is to use a polarization-selective coupling ({\em e.g.} elongated rectangular apertures) to read the cavity signal. For example, at 15 GHz a $2.8$mm-long section of  $4$mm$\times 0.6$mm rectangular waveguide preferentially transmits one polarization over the other by $36$dB (or a factor of $4.2\times 10^3$). 


Higher order (whispering gallery) TM$_{l1n}$ modes do have the right polarization to leak through the apertures.  Although they show up at frequencies higher than the desired TM$_{010}$ mode, the mode density can be very high, leading to confusion. These high order undesired TM$_{l1n}$ modes can be strongly suppressed with a multi-port coupling scheme that utilizes symmetry for phase cancellation. We propose to use a pre-amplification phase-matching summing network to achieve that goal. The binary summing network couples equally to the $2^N$ ports with the same propagation length, where $N$ is the number of levels. To understand how this guards against unwanted coupling, imagine in the reciprocal picture in which a signal is launched from the main branch, only the desired TM$_{010}$ mode will be excited, while other modes are suppressed by phase cancellation. With sufficiently large number of ports, the only modes (``grating lobes'') that can still add coherently to produce a signal at the main output branch necessarily will have much higher resonant frequencies. 

\begin{figure}[b]
\centering 

\includegraphics[width=3.8in]{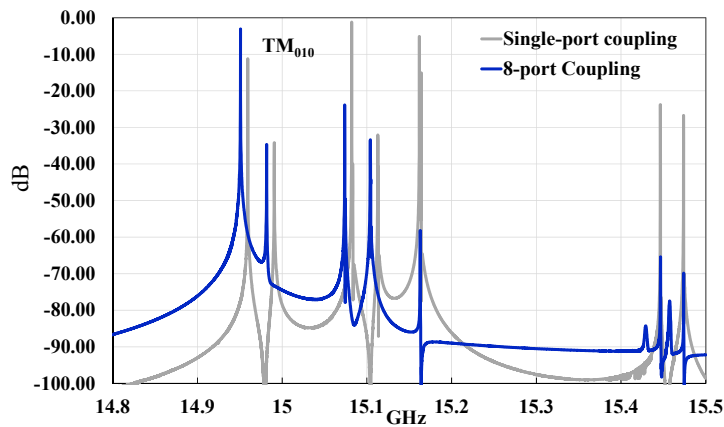}
\caption{Multi-port coupling suppresses spurious resonances in a large volume-cavity. FEA-calculated $S$-parameters for two coupling schemes are shown. In the first model, one input (port 0) and one output (port 1) are placed on the side wall of a circular shell-cavity ($r=2$ cm, $h=2$ cm). The $S_{10}$ parameter (gray) shows many spurious resonances in addition to the desired TM$_{010}$ mode. In the second model, eight singly polarized narrow slits (ports $1-8$) have been symmetrically placed on the inner and outer walls of the cavity in addition to the input port. The sum of $S$-parameters $\sum_{i=1}^8S_{i0}$ (blue) shows a significant enhancement of the TM$_{010}$ mode compared to the spurious modes. In an actual cavity, the adding is achieved with a pre-amplification summing network. }
\end{figure}

Only an experimental demonstration can decide whether this idea actually works for very large cavities. With that caution in mind, we go ahead and calculate the $S$-parameters for an 8-port coupled, modestly sized circular shell cavity as a proof of principle. With 3-D COMSOL calculations it is easy to verify that the signals from the eight ports add coherently in phase with an expected increase in the amplitude of the desired TM$_{010}$ resonance.  On the other hand, the combination of polarization selectivity and out-of-phase cancellation suppress other resonant modes despite the same increase in the coupling aperture.  This leads to a relative enhancement of the TM$_{010}$ mode compared to the spurious modes (FIG.6). It is important to note that the phase properties do not change fundamentally during tuning or linear superposition (``crossing'') of modes.

Another important function of this multi-port coupling scheme is to monitor against field localization. Fabrication defects lead to degeneracy breaking and undesired mode localization. For example, if the distance between the inner and outer walls is slightly larger at the top, a single TM$_{010}$ resonance will splits into two modes, localized at the top and bottom halves respectively. A single pick-up antenna located near the top of the cavity would only capture half of the active volume, and miss most of the sensitivity provided by the other new resonance now at a slightly higher frequency. This effectively reduces the active volume (form factor) of an axion haloscope.

The multi-port coupling scheme can mitigate that. Before the summing tree is connected, the off-diagonal scattering parameters of the $2^N$ ports can be measured. At the resonant frequency the $S_{ij}$ should all be identical both in amplitude and in {\em phase}. During the cavity assembly and testing stage, these phase-sensitive measurements of scattering parameters should be complementary to the bead-mapping in diagnosing fabrication errors and misalignments. Even if mode localization persists during operation, the spatially distributed ports can pick up all the localized modes at the fine structure frequencies so that the speed loss would be less severe than having a single coupling port.  


We should recall that the cold summing network is located outside the resonator. Therefore, the loss in the network represents a slight signal attenuation and does not degrade the $Q$-factor.  From past experiences, this somewhat complicated sounding scheme is actually quite manageable up to about 100 GHz (3 mm) for $N\sim 5$, or having $32$ coupling ports. An example for the similar architecture is the LO distribution networks used in CMB interferometers. Also, a spatial content-controlling pre-detection summing tree has been adopted very successfully in the detectors used for the BICEP/Keck program up to 270 GHz \cite{bicep2det}.

It is useful to compare the proposed scheme to ``Configuration 3'' discussed in \cite{jeong18}, in which a pre-amplification summing network is used to combine signal from multiple small cavities. A perfectly fabricated large volume-cavity has fewer mechanical degrees of freedom to control during tuning than many small cavities.  However, the price to pay for potentially simpler tuning/phase matching is the presence of a forest of other modes in a single large-volume cavity.  A more comprehensive discussion, including tuning strategies and coupling optimization will be presented in a follow-up paper.  In the next section, some preliminary designs are presented to show that the unusual geometry allows magnetic coupling and tuning in ways similar to standard axion haloscopes. 



\section{Magnetic coupling and Frequency Tuning}


Achieving high magnetic field strength over a large volume is one of the biggest challenges in an axion experiment. Therefore, a very appealing property of the new thin-shell geometry is that it can be placed in a conventional solenoid magnet with uniform field. For example, the ADMX main cavity space can readily house a circular thin-shell cavity that resonates at higher frequencies. This will be explored further in Section 5. 

On the other hand, if circular thin-shell cavities can be built with even larger diameters, they are most effectively coupled to pairs of nested Helmholtz coils. One can arrive at this configuration by moving half of the superconducting wires that would have been used to complete the solenoid magnet to inside the shell cavity. To first order, the magnetic field achieved in the cavity shell is comparable for the same amount of current and turns. However, an added benefit is that the counter-flowing current in the inner coils significantly reduces the field strength in the central region. Away from the shell cavity/coils, the field strength drops off quickly. This will ease the engineering effort associated with quench prevention.  The disadvantage of doing that is the magnets must now reside in the receiver cryostat and have to be custom-made.   


Another important aspect of an axion haloscope is frequency-tuning. 
The proposed cavities are similar to the single-rod cavity in HAYSTAC and can be tuned in a similar manner. To achieve sufficient tuning in the circular thin-shell cavity, we can split the top/bottom rings at the center circles and to create a $\sim \lambda/10$ ``tuning gap'' to work with. The inner wall assembly then can be moved laterally by {\em e.g.} a piezo nanopositioner, emulating the rotation of a single off-axis tuning rod. Alternatively, the inner tube can be physically deformed by an actuator into an ellipse, emulating the effects of two offset tuning rods. 

\begin{figure}[b]
\centering 

\includegraphics[width=4.8in]{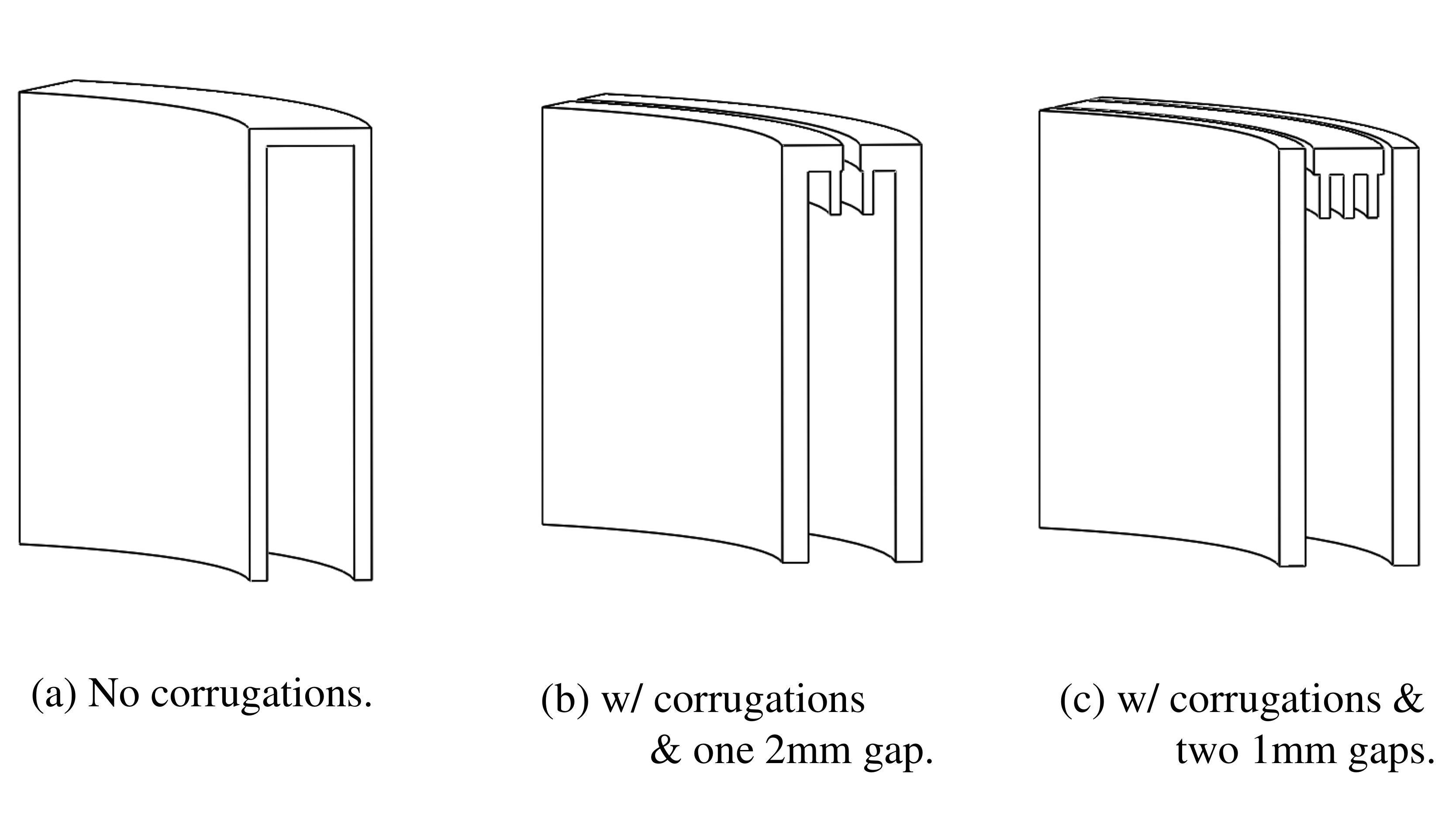}
\caption{\label{fig:epsart1} Sub-wavelength corrugations taper $E_z$ at the boundaries and confine the high-$Q$ eigenmodes even when these cavities have gaps (see FIG. 2). These circular gaps in the top/bottom rings separate the cavity into disjoint pieces that enable frequency tuning. With one gap ({\em center}) the inner tube can be displaced laterally or deformed; with two gaps ({\em right}) the top ring can be lower to reduce the height of the resonator. 
}
\end{figure}

There is a key implementation complication.  Before the split, the TM$_{010}$ mode induces zero surface current at the fault lines due to symmetry (FIG. 7a). However, if the cavity is simply split into two disjointed halves, the surface current on each side of the gaps can flow out vertically, effectively breaking the resonator.  Fortunately, this problem can be eliminated by adding azimuthal quarter-wave deep corrugations to the top and bottom rings, since such structure forces $E_\perp$ (and therefore voltage build-up) to vanish at the boundaries. 

However, as shown by these full 3D calculations, these corrugations kill off the $E$ field and the current at the top/bottom, and allow (sub-wavelength) gaps to exist for frequency tuning (FIG. 7b). The tuning range achievable by lateral displacements of the inner assembly of a circular shell cavity is shown in FIG. 8. Similar to the HAYSTAC/ADMX cavities, the broken azimuthal symmetry leads to mode localization and volume reduction \cite{hagmann90, lyapustin}.  On the other hand, FEAs show that the quality factor stays approximately unchanged throughout the tuning range.  

\begin{figure}[t]
\centering 
\includegraphics[width=3.5in]{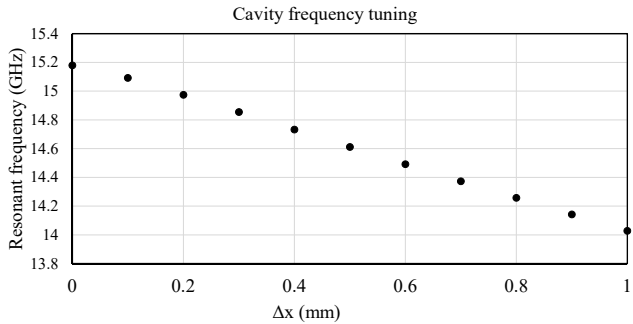}
\caption{Laterally displacing the inner assembly of a corrugated circular shell-cavity achieves a similar tuning range as a single rod-cylindrical cavity. As discussed in the main text, corrugations at the top/bottom of the cavity contain the field in the disjointed cavity (FIG. 7b). The results have been obtained using full 3D FEAs based on COMSOL-RF.}
\end{figure}

The newly created $z-$dependence by the corrugations also allows frequency tuning by changing the vertical configuration. This can be done either by moving the top (now corrugated) ``lid'' down, or by inserting a sapphire slab into the cavity space \cite{hagmann90}. Again, since the surface current already drops to zero near the corrugations, the gaps that are needed to mechanically realize the motion do not disrupt the cavity modes (FIG. 7c). Using Equation (1) (which is also appropriate for HE$_{11}$ modes for oversized apertures), we see that $10\%$ frequency tuning is achieved by reducing the cavity height from $5\lambda$ to $2\lambda$. This tuning scheme leads to disproportional reduction in volume, which is undesirable. However, mode localization mentioned above is avoided because azimuthal symmetry is completely maintained during tuning. Another advantage is that this scheme can work for the brain type cavity, which offers maximal increase in the volume and hence the scan rate albeit with a limited tuning range. Other non-symmetry-breaking tuning options are being studied and will be published in a follow-up work. 

\vspace{1cm}

\section{Projected Sensitivity and Summary}

Following the standard practice in CMB research, we will use actual achieved performance, instead of idealized first-principle calculations, to project the scan rate of a benchmark haloscope program based on circular thin-shell cavities. 

Recently, HAYSTAC experiment reported a limit on axion coupling strength at a factor of 2.7 above the benchmark KSVZ model \cite{marsh,graham} in the frequency range of 5.6 and 5.8 GHz. Obtained with a total of 240 days of data taking \cite{haystac_result}, this corresponds to a scan rate of 18 MHz/year for reaching KSVZ and will serve as the baseline for the scaling. 

The hypothesized pathfinder haloscope program will use a series of circular thin-shell cavities covering the frequency range between 5 and 10 GHz, each filling up to the inner diameter and the height of the {\em ADMX} cavity \cite{lyapustin}. The exact number of the shells depends on the achieved frequency tuning range.  At 5 GHz, this thin-shell cavity will have a factor of 15 larger active volume (including the form factor reduction from corrugations), or 225 times the scan rate of HAYSTAC, if everything else is held the same. At $\nu>5$ GHz, the higher axion mass leads to an enhanced scan rate $\propto \nu^2$. However, this scan rate enhancement will be hit by a factor of $V^2\propto \nu^{-2}$ from the shell width reduction ($w= \lambda/2$), and another factor of $\nu^{-2/3}$ from the anomalous skin effect that reduces $Q$. Finally, there is the degradation in the scan rate from the amplifier standard quantum limit, incurring another factor of $\nu^{-2}$ . 

Taken together, the scaled scan rate is approximately $d\nu/dt=4.0(\nu/5.7{\rm GHz})^{-8/3}$ GHz/year. The integral $\int_5^{10}(d\nu/dt)^{-1} d\nu$ equals 2.8 years. In other words, it takes about three years to cover the frequency between 5 and 10 GHz to KSVZ sensitivity using quantum-limited amplifiers as readout, not including warm-up/cool-down time required for swapping shell cavities and readout systems.  Assuming a $15\%$ tuning range, the number of required shells to cover an octave of frequency is five: $5-5.75$ GHz, $5.75-6.61$ GHz, $7.60-8.75$ GHz, and $8.75-10.0$ GHz. Over several years of integration, the down time would be insignificant with only five sets of shells. Although in practice, more shells might be needed to cover the frequency range missing due to mode crossing. 

This simple estimate shows the great potential of thin-shell cavities for axion searches even at the low frequency end of cm wavelength. Serious R\& D efforts into fabrication and coupling issues will be very well justified.

\acknowledgments

The author acknowledges the support of a KIPAC Innovation Grant and useful discussions with Zeesh Ahmed, Karl van Bibber, Jamie Bock, Ari Cukierman, Peter Graham, Kent Irwin, Jeff McMahon, David Schuster, and Sami Tantawi.

\nocite{*}

\bibliographystyle{JHEP}
\bibliography{ms.bib}









\end{document}